\documentclass[traditabstract,letter]{aa} % for the abstract without structuration 
\usepackage{graphicx}
\usepackage{txfonts}
\newcommand{\Rlag}{R_{\rm \tau_K}}

\newcommand{\Rring}{R_{\rm ring}}

\newcommand{\Vsys}{V^2_{\rm sys}}

\usepackage{natbib}
\begin{document}
\title{Exploring the inner region of Type 1 AGNs with the Keck
  interferometer}

   %\subtitle{}

\author{Makoto Kishimoto$^1$,
       Sebastian F. H\"onig$^1$,
       Robert Antonucci$^2$, 
       Takayuki Kotani$^3$, 
       Richard Barvainis$^4$, 
       Konrad R.~W.~Tristram$^1$,
       \and
       Gerd Weigelt$^1$
%        et al.
%       \fnmsep\thanks{}
       }

\institute{
  $^1$Max-Planck-Institut f\"ur Radioastronomie, Auf dem H\"ugel 69,
  53121 Bonn, Germany; \email{mk@mpifr-bonn.mpg.de}.  $^2$Physics
  Department, University of California, Santa Barbara, 93106,
  USA. $^3$ISAS, JAXA, 3-1-1 Yoshinodai, Sagamihara, Kanagawa, 
  229-8510 Japan. $^4$National
  Science Foundation, 4301 Wilson Boulevard, Arlington, 
  VA 22230, USA.
%      \and
%          \\
%          \email{}
%          \thanks{}
          }

   \date{Submitted 20 October 2009 / Accepted 3 November 2009}

\authorrunning{Kishimoto et al.}

\titlerunning{Exploring the inner region of Type 1 AGNs}

% \abstract{}{}{}{}{} 
% 5 {} token are mandatory
 
\abstract{ 

  The exploration of extragalactic objects with long-baseline
  interferometers in the near-infrared has been very limited. Here we
  report successful observations with the Keck interferometer at
  K-band (2.2 $\mu$m) for four Type 1 AGNs, namely NGC4151, Mrk231,
  NGC4051, and the QSO IRAS13349+2438 at $z$=0.108.  For the latter
  three objects, these are the first long-baseline interferometric
  measurements in the infrared. We detect high visibilities ($V^2
  \sim$ 0.8$-$0.9) for all the four objects, including NGC4151 for
  which we confirm the high $V^2$ level measured by Swain et
  al.~(2003). We marginally detect a decrease of $V^2$ with increasing
  baseline lengths for NGC4151, although over a very limited range,
  where the decrease and absolute $V^2$ are well fitted with a ring
  model of radius 0.45$\pm$0.04~mas (0.039$\pm$0.003~pc).  Strikingly,
  this matches independent radius measurements from optical--infrared
  reverberations that are thought to be probing the dust sublimation
  radius.  We also show that the effective radius of the other
  objects, obtained from the same ring model, is either roughly equal
  to or slightly larger than the reverberation radius as a function of
  AGN luminosity. This suggests that we are indeed partially resolving
  the dust sublimation region. The ratio of the effective ring radius
  to the reverberation radius might also give us an approximate probe
  for the radial structure of the inner accreting material in each
  object. This should be scrutinized with further observations.

}


\maketitle
%
%________________________________________________________________

\section{Introduction}

The exploration of extragalactic objects, or in particular, Active
Galactic Nuclei (AGNs), with long-baseline interferometers in the
near-infrared (near-IR) has been very limited. While the brightest
Type 1 AGN NGC4151 and Type 2 AGN NGC1068 have been observed by
\cite{Swain03} and \cite{Wittkowski04}, respectively, further
exploration has been hampered mainly by technical difficulties.  Here
we report successful observations of four Type 1 AGNs with the Keck
interferometer (KI) in the near-IR (K-band 2.2 $\mu$m).  Type 1 AGNs
are thought to give us a direct view of the innermost region of the
putative dust torus as well as the central accretion disk, where the
interesting effect of the latter should also be evaluated carefully.

\section{Keck interferometry}

\subsection{Observations and data reduction}\label{sec_kiobs}

We observed four AGNs listed in Table~\ref{tab_obs} and associated
calibrators with the Keck Interferometer (KI; \citealt{Colavita03}) on
2009 May 15 (UT).  These four targets were chosen based on their
bright optical magnitudes measured from the pre-imaging data obtained
in April 2009 at Tiki Observatory (French Polynesia) and Silver Spring
Observatory (USA) by N.~Teamo, J.~C.~Pelle, and K.~Levin.

The KI combines the two beams from the two Keck 10~m telescopes which
are separated by 85~m along the direction 38\degr\ east of
north. Adaptive Optics correction was implemented at each telescope,
locking on the nucleus in the visible wavelengths.  The data
were obtained with a fringe tracker rate of 200 Hz operated at K-band,
while the angle-tracking was performed at H-band. The data were first
reduced with {\sf Kvis}\footnotemark[1] to produce raw squared visibility
($V^2$) data averaged over blocks of 5 sec each.

Then those blocks with a phase RMS jitter larger than 0.8 radian
were excluded.  For a given visit of each object, the blocks with an
AO wavefront sensor flux smaller than the median by 20\% or more were
also excluded. The rejected blocks were $\sim$ 6\% of all the blocks,
and they were generally outliers in the visibility measurements for a
given visit, and quite often were of low fractional fringe-lock time.
Then the wide-band side of the data were further reduced using {\sf
  wbCalib}\footnotemark[2] with the correction for the flux ratio between two
telescope beams and the correction for a flux bias (slight systematic
decrease of the KI's system visibility for lower injected
flux\footnotemark[3]). The jitter correction was applied with a
coefficient 0.04 \citep{Colavita99pasp}.  Then the blocks were
averaged into scans over a few minutes each, with its error estimated
as a standard deviation within a scan.

\footnotetext[1]{\tiny{http://nexsci.caltech.edu/software/KISupport/v2/V2reductionGuide.pdf}}

\footnotetext[2]{\tiny{http://nexsci.caltech.edu/software/V2calib/wbCalib/index.html}}

\footnotetext[3]{\tiny{http://nexsci.caltech.edu/software/KISupport/dataMemos/index.shtml}}

\begin{figure}
\centering \includegraphics[width=9cm]{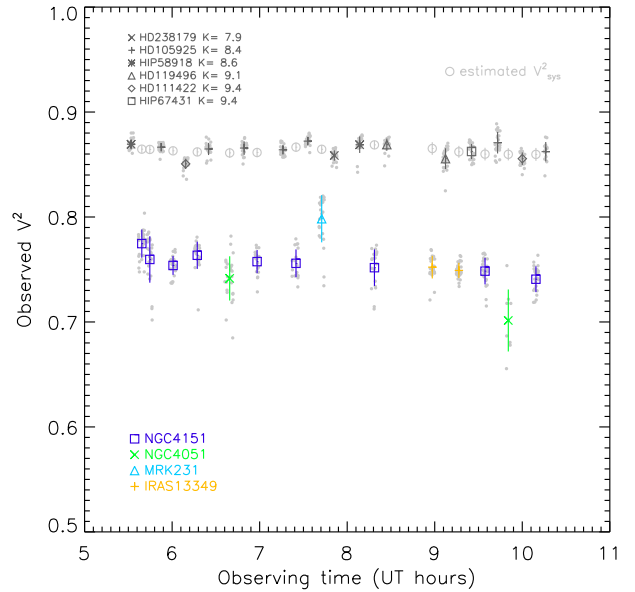}
\caption{Observed $V^2$ plotted against observing time. Gray dots are
  individual measurements for blocks of 5 sec each. Gray circles are
  the estimated system $V^2$ at the time of target observations.}
\label{fig_rawvis}
\end{figure}

Fig.\ref{fig_rawvis} shows observed visibilities of all the targets
and calibrators (after the corrections above), plotted
against observing time.  All the six calibrators used are expected to
be unresolved by the KI at K-band ($V^2 \ge 0.999$).  Overall, the
system visibility ($\Vsys$), as measured by these calibrators,
was quite stable over the night.  The calibrators span over a
relatively wide range of brightness (see the legend in
Fig.\ref{fig_rawvis}), and one of them (HD111422) had approximately
the same injected flux counts as those of NGC4151 and Mrk231.  Based
on the corrected visibilities of these calibrators shown in
Fig.\ref{fig_rawvis}, the flux bias seems to have been taken out quite
well, although there might still be some systematics left.  The
visibilities of the three faint calibrators ($K>9.1$) tend to be
slightly smaller than those for the other brighter ones, with the
difference of the means of the former and latter being $\sim$0.9\% of
the means.  Therefore we assign 0.01 as a possible systematic
uncertainty in system visibility estimations.

Note that for the faintest target NGC4051, the flux bias correction
was effectively used with a slight extrapolation (by $\sim$1 magnitude
in K-band\footnotemark[3]).  This should be checked with future
fainter calibrator observations.

% (but note that at least we did not see any obvious evidence for a
% visibility decrease as a function of fluctuating level of injected
% flux for each target).

\begin{figure}
\centering \includegraphics[width=9cm]{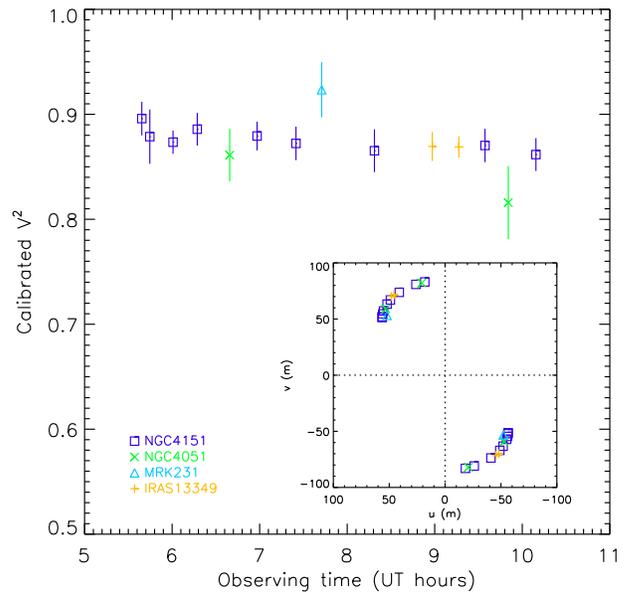}
\caption{Calibrated $V^2$ plotted against observing time. The inset
  shows the sampled uv points for each target (north to the top,
  east to the left).}
\label{fig_calvis_uv}
\end{figure}

For each target measurement, $\Vsys$ was estimated from these
calibrator observations with {\sf wbCalib} using its time
and sky proximity weighting scheme, yielding $\Vsys$ indicated as
gray circles in Fig.\ref{fig_rawvis}.  The final calibrated
visibilities are shown in Fig.\ref{fig_calvis_uv}, with the sampled $uv$
points shown in the inset of Fig.\ref{fig_calvis_uv}.

% The angle-tracking 80Hz or 40Hz.

% differential phase spectra

\begin{table*}
%{\tiny

\caption[]{Properties of our targets and summary of the results of our KI observations on 15 May 2009 (UT).}
{\tiny
\begin{tabular}{lllccccccccccccc}
\hline
name   & $z^a$  & scale$^b$      & $E_{B-V}$$^c$ & $B_{p}$$^d$ & PA & $V^2$ ($^e$)& \multicolumn{2}{c}{$\Rring$$^f$}  & $\Rlag$$^g$   & $A_{V}$   \\ 
\cline{8-9}
       & corr.& (pc mas$^{-1}$)  &  (mag)  & (m)   & (\degr)   &             & (mas)        & (pc)         & (pc)     & (mag) \\
\hline
NGC4151        & 0.00414 & 0.086 & 0.028 & 85.0 & 12.3 & 0.862$\pm$0.018 & 0.45$\pm$0.04 & 0.039$\pm$0.003 & $\sim$0.044$\pm$0.011 & \\ 
NGC4051        & 0.00309 & 0.064 & 0.013 & 79.4 & 41.9 & 0.861$\pm$0.026 & 0.51$\pm$0.05 & 0.032$\pm$0.003 & $\sim$0.011$\pm$0.004 & \\
MRK231         & 0.0427  & 0.84  & 0.010 & 74.4 & 44.3 & 0.923$\pm$0.028 & 0.38$\pm$0.07 & 0.32$\pm$0.06   &                       & 1.3$^h$\\
IRAS13349+2438 & 0.109   & 2.0   & 0.012 & 85.0 & 34.5 & 0.869$\pm$0.016 & 0.44$\pm$0.03 & 0.88$\pm$0.05   &                       & 0.93$^i$\\
\hline
\end{tabular}
\\
$^a$ CMB corrected value from NED. 
$^b$ $H_0=70$ km s$^{-1}$ Mpc$^{-1}$, $\Omega_{\rm m}=0.3$, and $\Omega_{\Lambda}=0.7$.
$^c$ Galactic reddening from \cite{Schlegel98}. $^d$ Projected baseline lengths. 
$^e$ For objects with multiple measurements, $V^2$ data shown are at
the longest baseline length, except for NGC4051 where the one with a
smaller error is shown. $^f$ Thin-ring radius (best-fit values if multiple data are available). 
$^g$ Mean and standard deviation of the reverberation measurements
over 2001 to 2006 from \cite{Koshida09} for NGC4151, and over 2001 to
2003 from \cite{Suganuma06} for NGC4051.
% Minezaki et al. 0.040 pc for NGC4151
% Suganuma et al. 0.009 - 0.015 pc for NGC4051
$^h$ \cite{Lacy82}.
$^i$ \cite{Wills92}.
}

%}
\label{tab_obs}
\end{table*}

\subsection{Results}

\begin{figure}
\centering \includegraphics[width=9cm]{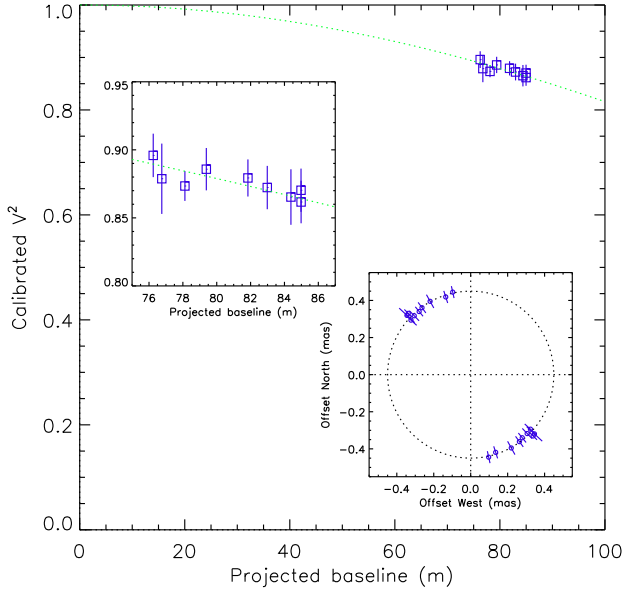}
\caption{Calibrated $V^2$ for NGC4151 as a function of projected
  baselines, enlarged in the left inset. The dotted line shows the
  best-fit visibility curve with a ring model of radius 0.45 mas.  In
  the right inset, ring radii are plotted along the PA of each
  projected baseline. Note that the correction for the accretion disk
  and host galaxy contributions is not incorporated in this figure,
  and it does not change the ring radius significantly (see
  Table~\ref{tab_corr}).}
\label{fig_ngc4151}
\end{figure}

Fig.\ref{fig_ngc4151} shows the final calibrated visibilities for
NGC4151 as a function of projected baseline length, enlarged in the
left panel of the insets.  We confirm the visibility level observed by
\cite{Swain03}.  The covered range of projected baselines is very
limited (note also that the shortest possible for NGC4151 is $\sim$70
m due to the KI's delay line restriction).  We see, however, a
marginal decrease of visibility over the increasing baseline.  With
the Spearman's rank correlation coefficient analysis, the confidence
level is 98.4\%, or 2.4 $\sigma$.  The decrease and absolute level of
visibilities are well fitted with a simple thin ring model (i.e.
inner radius equal to outer) of radius $0.45\pm0.04$ mas
($0.039\pm0.003$ pc; the error accounts for the systematic uncertainty
in $\Vsys$; Sec.\ref{sec_kiobs}). If we convert each of the visibility
measurements into a ring radius and plot it as a function of the PA of
the projected baseline, we obtain the right panel of the insets in
Fig.\ref{fig_ngc4151}. Over the PA range covered, from $\sim$10\degr\
to $\sim$50\degr, we do not seem to see PA dependence of the radius.
For all the other targets, the calibrated visibilities are shown in
Table~\ref{tab_obs} together with baseline information and deduced
ring radii.

\section{UKIRT imaging and KI data corrections}\label{sec_ukirt}

In order to obtain quasi-simultaneous flux measurements for the
nuclear point-source in the four Type 1 AGNs (which are variable),
their images were obtained with WFCAM on UKIRT in five broad-band
filters (Table~\ref{tab_corr}) on 2009 June 17 (UT) under the UKIRT
service programme.  The seeing was $\sim$1 arcsec.  The
pipe-line-reduced data were obtained through the WFCAM Science
Archive.  The wide field of view of the WFCAM gave simultaneous
measurements of PSF stars in each AGN field, while a 2x2 micro
stepping gave a good image sampling with an effective pixel size of
0.2 arcsec. We implemented two-dimensional (2D) fits for each image to
accurately separate the PSF component from the underlying host galaxy,
following the same procedure as described by \cite{Kishimoto07}.

Table~\ref{tab_corr} lists the measured flux of the nuclear PSF
component. The uncertainty of our nuclear PSF flux measurements is
estimated as $\sim$5\%, based on the residual fluxes after the fits
and the flux calibration uncertainty.  In the K-band images of the two
brighter objects, namely NGC4151 and Mrk231, the central several
pixels seemed affected by non-linearity or saturation, so we
implemented the 2D fits by masking the central $\sim$0.4 arcsec radius
region. We estimate that the nuclear PSF flux is recovered within the
same uncertainty of 5\%, based on the results from the same masked
fits on the other unsaturated images.  Using the results of the
PSF--host decomposition, we also estimated the host galaxy flux
fraction within the field-of-view of the KI which is $\sim$50 mas at
K-band (FWHM; set by a single-mode fiber for the fringe tracker).  The
results are stated in Table~\ref{tab_corr}. The obtained small values
show that the host galaxy contribution is only a very small part of
the observed visibility departure from unity.

\begin{figure}
\centering \includegraphics[width=8cm]{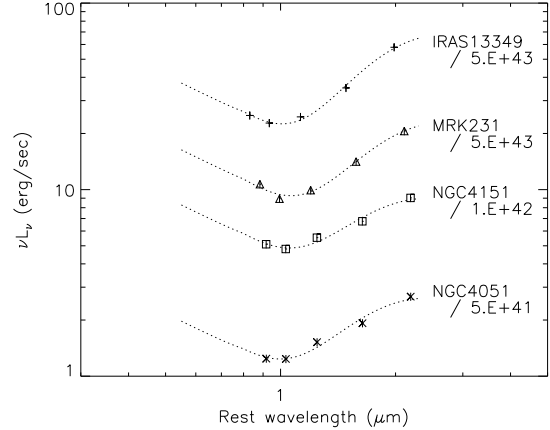}
\caption{Flux of the nuclear PSF component in WFCAM images derived
  from 2D fits. Fitted SEDs are shown in dotted lines (see text).}
\label{fig_sed}
\end{figure}

Fig.\ref{fig_sed} shows the resulting spectral energy distribution
(SED) of the PSF component in each target, after the correction for
Galactic reddening.  We also corrected for the reddening in the host
galaxy for the objects which show large Balmer decrements in broad
emission lines (Table~\ref{tab_corr}).  Assuming that the PSF flux
originates from the hot dust thermal emission nearly at the
sublimation temperature and from the central accretion disk (AD;
thought to be directly seen in Type 1 inclinations), we estimate the
flux fraction at K-band from the latter AD component. Here we fit the
SED with a power-law spectrum of the form $f_{\nu} \propto \nu^{+1/3}$
for the AD, plus a spectrum of a black-body form for the dust (the
best-fit temperature was $\sim$1300$-$1500 K).  The AD flux fraction
at K-band is estimated to be as small as $\sim$0.2
(Table~\ref{tab_corr}), in agreement with the results by
\cite{Kishimoto07}.  This suggests that the high visibilities observed
are not due to the unresolved AD, as opposed to the preferred
interpretation by \cite{Swain03} for NGC4151.  The assumed near-IR AD
spectral form is based on the recent study of near-IR polarized flux
spectra \citep{Kishimoto08}, but also on various studies on AD
spectral shapes in the optical/UV (summarized in Fig.2 of
\citealt{Kishimoto08}). By assigning the uncertainty in the AD near-IR
spectral index as 0.3, we also estimated the uncertainty of the K-band
AD flux fraction (Table~\ref{tab_corr}).

Finally, we corrected the observed visibilities for the host
galaxy and the AD contributions, where the latter is assumed to remain
unresolved by the KI. The corrected $V^2$ as well as corresponding
thin ring radii are listed in Table~\ref{tab_corr}.

\begin{table*}
%{\tiny
\caption[]{Point-source flux from UKIRT imaging data and the KI results corrected for host galaxy and unresolved AD component.} 
{\tiny
\begin{tabular}{llcccccccccccccccccccc}
\hline
name    & \multicolumn{5}{c}{flux (mJy)} & & \multicolumn{3}{c}{magnitude} & host$^a$  & AD    & $V^2$ corrected & $\Rring$ corr.   & $\Rlag$ fit$^c$\\ 
\cline{2-6}  \cline{8-10}
        & Z & Y & J & H & K              & & J & H & K                     & (\%)     & fraction$^b$ &                 & (pc)          & (pc) \\ 
\hline
NGC4151        & 39.7 & 42.4 & 59.4 & 96.3 & 173. && 11.0 & 10.1 &  8.9 & 0.3$\pm$0.1 & 0.17$\pm$0.06 & 0.841$\pm$0.024 & 0.041$\pm$0.004 & 0.033 \\ 
NGC4051        & 8.88 & 9.98 & 14.9 & 24.8 & 46.1 && 12.6 & 11.6 & 10.4 & 1.5$\pm$1.0 & 0.14$\pm$0.05 & 0.870$\pm$0.038 & 0.032$\pm$0.005 & 0.011 \\
Mrk231         & 21.0 & 22.0 & 33.5 & 71.6 & 152. && 11.7 & 10.4 &  9.1 & 0.5$\pm$0.1 & 0.15$\pm$0.05 & 0.921$\pm$0.033 & 0.33$\pm$0.07   & 0.33  \\
IRAS13349      & 7.75 & 8.71 & 12.6 & 26.3 & 62.3 && 12.7 & 11.5 & 10.0 & 0.4$\pm$0.1 & 0.14$\pm$0.04 & 0.856$\pm$0.021 & 0.92$\pm$0.06   & 0.49  \\
\hline
\end{tabular}
\\
$^a$ Host galaxy flux fraction at K-band estimated for the KI's 50 mas
FOV in the AO-corrected images. $^b$ AD flux fraction of the point source at K-band.
$^c$ $\Rlag$ from UV luminosity using the fit by \cite{Suganuma06}.
It has an uncertainty of a factor of $\sim$1.5 based on the scatter
of the fit (see text).
}
%}
\label{tab_corr}
\end{table*}

\section{Interpretations and discussions}

\begin{figure}
\centering \includegraphics[width=8cm]{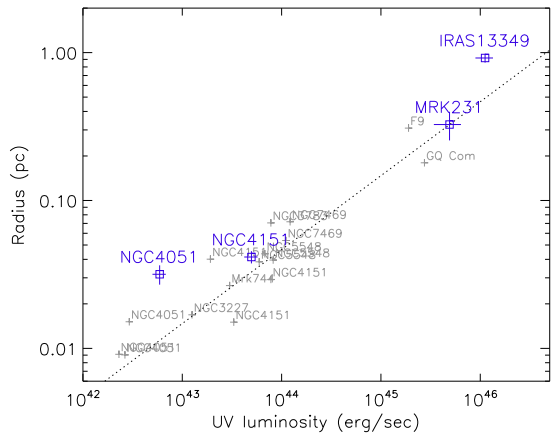}
\caption{Corrected ring radius derived for each KI target
  (squares), plotted against UV luminosity, or a scaled V-band
  luminosity (extrapolated from the WFCAM Z-band flux; see text). Also
  shown in gray plus signs are the reverberation radii against the
  same scaled V-band luminosity \citep[][and references
  therein]{Suganuma06} and their fit (dotted line).}
\label{fig_radius_pc}
\end{figure}

We interpret here the high visibilities observed for all the four
objects as an indication of partially resolving the inner brightness
distribution of dust thermal emission.

As we discussed in Sec.\ref{sec_ukirt}, the AD flux fraction at K-band
is estimated to be small, as long as the assumed power-law AD spectrum
in the near-IR, smoothly continuing from the optical, is at least
roughly correct.  In this case, the K-band emission is dominated by
the dust emission, and it is reasonable to convert the observed
visibility to a thin ring radius to obtain an approximate effective
radius of the dust brightness distribution for each object. (We have
corrected $V^2$ for the unresolved AD contribution, but the correction
is quite small; Table~\ref{tab_corr}.) The derived ring radii are
plotted in parsec in Fig.\ref{fig_radius_pc} against UV luminosity
$L$, here defined as a scaled V-band luminosity of $6\ \nu f_{\nu}
(V)$ \citep{Kishimoto07}. The V-band flux is extrapolated from the
fitted flux at 0.8 $\mu$m (Fig.\ref{fig_sed}) assuming an AD spectral
shape of $f_{\nu} \propto \nu^{0}$ (based on spectral index studies
referred to above).

We can directly compare these ring radii with another type of
independent radius measurements $\Rlag$, namely the light traveling
distance for the time lag of the K-band flux variation from the
UV/optical variation \citep{Suganuma06}.  These reverberation radii
are also plotted against the same scaled V-band luminosity in
Fig.\ref{fig_radius_pc}.  They are known to be approximately
proportional to $L^{1/2}$ (\citealt{Suganuma06}; the dotted line in
Fig.\ref{fig_radius_pc} shows their fit), and are likely to be probing
the dust sublimation radius in each object.  We first see that
$\Rring$ is roughly comparable to $\Rlag$ for all the four objects,
and thus $\Rring$ is roughly scaling also with $L^{1/2}$. This
approximate match suggests that the KI data are indeed partially
resolving the dust sublimation region.

With a closer look at Fig.\ref{fig_radius_pc}, we see that $\Rring$ is
either roughly equal to or slightly larger than $\Rlag$
(i.e. $\Rring/\Rlag \gtrsim 1$, up to a factor of a few), though we
have only four objects.  This could be understood if $\Rlag$ is
tracing a radius close to the innermost boundary radius of the dust
distribution.  It is known that the cross-correlation lag tends to
trace an inner radius of the responding particles' distribution when
the lag is determined from the peak in the cross-correlation function
\citep[e.g.][and references therein]{Koratkar91}, as is the case for
the data used in the fit by Suganuma et al.  On the other hand,
$\Rring$ is an effective, average radius over the radial dust
brightness distribution in the K-band.  When the radial distribution
is very steep and compact, the ratio $\Rring/\Rlag$ would become very
close to unity (such as seen in NGC4151 and Mrk231), while for a
flatter, more extended distribution, $\Rring /\Rlag$ would show a
larger departure from unity.

If our interpretation above is correct, the KI data would conversely
support the dust sublimation radius as probed by the reverberation
measurements that is smaller by a factor of about three than that
inferred for typical ISM-size Graphite grains \citep[][0.05 $\mu$m
radius]{Barvainis87} for a given $L$ \citep{Kishimoto07}.  The small
sublimation radius could be due to the possible dominance of large
grains in the innermost region, since they can sustain at a much
closer distance to the illuminating source for a given sublimation
temperature.  Alternatively, it could be due to an anisotropy of the
AD radiation (see \citealt{Kishimoto07} for more details).

In Fig.\ref{fig_radius_pc}, the reverberation and ring radii are shown
essentially as a function of an instantaneous $L$ at the time of each
corresponding radius measurement. \cite{Koshida09}, however, recently
have shown that $\Rlag$ is not exactly scaling with the instantaneous
$L$ varying in a given object.  It might be that $\Rlag$ tends to give
a dust sublimation radius that corresponds to a relatively long-term
average of $L$. On the other hand, $\Rlag$ does show the $L^{1/2}$
proportionality over a {\it sample} of objects. Thus, when we compare
$\Rlag$ and $\Rring$, unless simultaneous measurements exist, we would
have to allow for the uncertainty in $\Rlag$, as a function of
instantaneous $L$, being the scatter in the $L^{1/2}$ fit ($\sim$0.17
dex).

% We also note that the radial structure issue above is better addressed
% together with mid-IR interferometric data, as we have demonstrated in
% \cite{Kishimoto09}.  We have mid-IR data for several Type~1s,
% including simultaneous measurements for two of the KI sample presented
% here (NGC4151 and IRAS13349+2438). We will present analyses of these
% data with the present KI data in a forthcoming paper.

If the AD spectrum does not have a power-law shape but rather has some
red turn-over in the near-IR, though the near-IR polarized flux
spectrum argues against this \citep{Kishimoto08}, the AD flux fraction
at K-band would become higher than we estimated here.  However, even
if the flux fraction is as large as 0.5, the corrected ring radius
would become larger than stated in Table~\ref{tab_corr} by a factor of
only $\sim$1.3, resulting in no qualitative change in our discussion.
Future near-IR interferometry with much longer baselines can
conclusively confirm that the visibility is decreasing as we inferred
from the present KI data.  We plan to advance our exploration with
further interferometric measurements in the infrared.

%\section{Conclusions}

\begin{acknowledgements}
%% This research used the facilities
%% of the Canadian Astronomy Data Centre operated by the National
%% Research Council of Canada with the support of the Canadian Space
%% Agency. 

%% This research is partly based on observations made with the European
%% Southern Observatory telescopes obtained from the ESO/ST-ECF Science
%% Archive Facility.  

%% This research has made use of the NASA/IPAC
%% Extragalactic Database (NED) which is operated by the Jet Propulsion
%% Laboratory, California Institute of Technology, under contract with
%% the National Aeronautics and Space Administration.

%% This research is partly based
%% on observations made with the NASA/ESA Hubble Space Telescope,
%% obtained from the Data Archive at the Space Telescope Science
%% Institute, which is operated by the Association of Universities for
%% Research in Astronomy, Inc., under NASA contract NAS 5-26555.

  The data presented herein were obtained at the W.M. Keck
  Observatory, which is operated as a scientific partnership among the
  California Institute of Technology, the University of California and
  the National Aeronautics and Space Administration. The Observatory
  was made possible by the generous financial support of the W.M. Keck
  Foundation.  We are grateful to all the staff members whose huge
  efforts made these Keck interferometer observations possible.  The
  United Kingdom Infrared Telescope is operated by the Joint Astronomy
  Centre on behalf of the Science and Technology Facilities Council of
  the U.K.  We thank N.~Teamo, J.~C.~Pelle and K.~Levin for kindly
  providing the pre-imaging data, and F. Millour for helpful
  discussions.  This work has made use of services produced by the
  NASA Exoplanet Science Institute at the California Institute of
  Technology.

\end{acknowledgements}

\end{document}